\documentclass[vecphys]{svmult}

\usepackage{makeidx}         
\usepackage{graphicx}        
\usepackage{multicol}        
\usepackage[bottom]{footmisc}

\makeindex             

\begin{document}

\newcommand{\be}{\begin{equation}}
\newcommand{\ee}{\end{equation}}
\newcommand{\nn}{\nonumber}
\newcommand{\bea}{\begin{eqnarray}}
\newcommand{\eea}{\end{eqnarray}}
\newcommand{\bfig}{\begin{figure}}
\newcommand{\efig}{\end{figure}}
\newcommand{\bc}{\begin{center}}
\newcommand{\ec}{\end{center}}
\def\ad{\dot{\alpha}}
\def\ov{\overline}
\def\hlf{\frac{1}{2}}
\def\qrt{\frac{1}{4}}
\def\as{\alpha_s}
\def\at{\alpha_t}
\def\ab{\alpha_b}
\def\sq2{\sqrt{2}}
\newcommand{\smallz}{{\scriptscriptstyle Z}} %
\newcommand{\mz}{m_\smallz}
\newcommand{\smallw}{{\scriptscriptstyle W}}
\newcommand{\mw}{m_\smallw} 
\newcommand{\gw}{\Gamma_{\smallw}} 
\newcommand{\sw}{s_{\smallw}} 
\newcommand{\cw}{c_{\smallw}} 
\newcommand{\sdw}{\sin^2\theta_{\smallw}} 
\newcommand{\cdw}{\cos^2\theta_{\smallw}} 
\newcommand{\sqw}{\sin^4\theta_{\smallw}} 
\newcommand{\smallh}{{\scriptscriptstyle H}}
\newcommand{\mh}{m_\smallh}
\newcommand{\mt}{m_t}
\newcommand{\wh}{w_\smallh}
\newcommand{\oa}{${\cal O}(\alpha)~$} 
\newcommand{\oab}{${\cal O}(\alpha)$} 
\def\th{t_\smallh}
\def\zh{z_\smallh}
\newcommand{\Mvariable}[1]{#1}
\newcommand{\Mfunction}[1]{#1}
\newcommand{\Muserfunction}[1]{#1}
\newcommand{\smartpap}{p\hskip-7pt\hbox{$^{^{(\!-\!)}}$}}

\title*{Electroweak corrections to the charged-current Drell-Yan process}
\author{
C.M. Carloni Calame \inst{1,2} \and
G. Montagna \inst{2,1} \and
O. Nicrosini \inst{1} \and
A. Vicini\inst{3}
}

\institute{
INFN, Sezione di Pavia, Via A. Bassi 6, Pavia (Italy) \and
Dipartimento di Fisica Nucleare e Teorica, Via A. Bassi 6, Pavia (Italy) \and
Dipartimento di Fisica and INFN-Sezione di Milano, Via Celoria 16, 20133 Milano (Italy)
}

%
%
\maketitle

{\bf Introduction.}
The production of a high-transverse-momentum lepton pair,
known as Drell-Yan (DY) process, 
plays an important role at hadron colliders,
such as the Fermilab Tevatron and the CERN LHC, thanks to the large
cross section 
and the clean signature of the final state, with at
least one high-transverse-momentum lepton to trigger on.
After the LEP and Tevatron precision measurement of the gauge boson
masses, the DY processes represent standard candles which can
be used for the detector calibration.
The production of gauge bosons, in association with jets, is an
important background to interesting physics channels, like the top
quark pair production.
The mass of the $W$ boson will be measured at the LHC from the
transverse mass distribution, but also from the ratio
$(d\sigma^W/dM_\perp)/(d\sigma^Z/dM_\perp)$,
with a foreseen final uncertainty of $\Delta\mw\sim 15$ MeV \cite{TDR}. 
The latter, combined with the 
improvement in the determination of the top quark mass
(foreseen with an accuracy of 1-2 GeV),
will put more stringent bounds in all the 
precision tests of the Standard Model.
The DY processes provide, both in neutral and charged current channels,
stringent constraints on the density functions which
describe the partonic content of the proton \cite{pdf}.
The important progress in the calculation of the QCD corrections
has reduced at the per cent level 
the residual theoretical uncertainties which affect the
DY cross sections; as a consequence, it has been proposed to use them
as a luminosity monitor of the collider \cite{DPZ,FM}.
The large mass tail of the invariant mass distribution represents an
important background 
to the search for new heavy gauge bosons
\cite{CDFnewgauge}.

\noindent
{\bf QCD calculations and generators.}
The accuracy in the determination of the theoretical cross-section has
greatly increased over the years.
Next-to-next-to-leading order (NNLO)
QCD corrections to the total cross section have been computed
in~\cite{HvNM}, but  differential distributions with the same accuracy
have been obtained only recently in~\cite{ADMP}.
A realistic phenomenological study and the data analysis require
the inclusion of the relevant radiative corrections 
and their implementation into Monte Carlo event generators,
in order to simulate all the experimental cuts 
and to allow, for instance, an accurate determination of the 
detector acceptances.
The Drell-Yan processes are included in the standard QCD Parton Shower
generators $\tt HERWIG$ and $\tt PYTHIA$~\cite{HERWIG,PYTHIA}.
Recently there have been important progresses to improve the QCD
radiation description to NLO, which has been implemented in 
the code $\tt MC@NLO$~\cite{MC@NLO}.
Another important issue is the good description of the 
intrinsic transverse momentum of the gauge boson, which can be
obtained by resumming up to all orders the contributions of the form
$\alpha_s\log(p_\perp^{\scriptscriptstyle W}/\mw)$.
The generator $\tt RESBOS$~\cite{BY}, used for data analysis at
Tevatron, includes these effects.

\noindent
{\bf Electroweak calculations and generators.}
The size of the NNLO QCD corrections and the improved stability of the
results against changes of the renormalization/factorization scales
raises the question of the relevance of the \oa electroweak (EW) radiative
corrections, which were computed, in the charged current channel,
first in the pole approximation~\cite{HW,BKW} and then fully
in~\cite{DK,BW,SANC}.
\begin{figure}
\centering
\includegraphics[height=35mm]{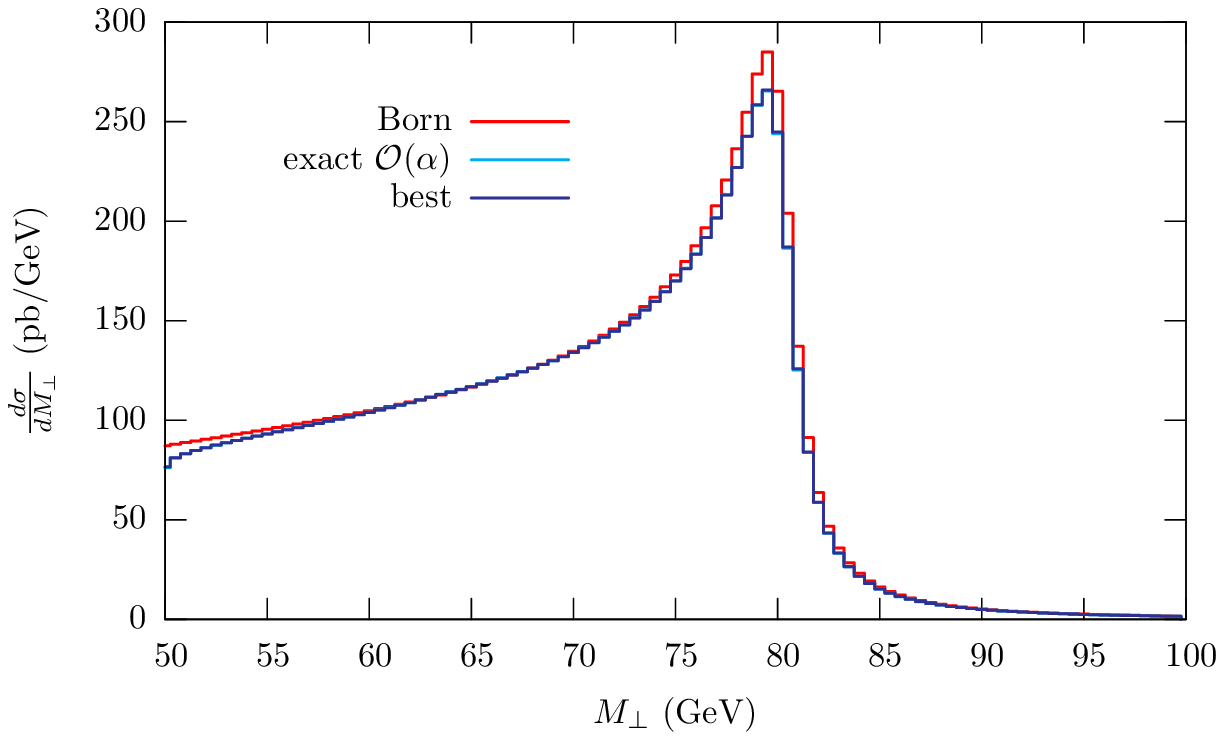}~
\includegraphics[height=35mm]{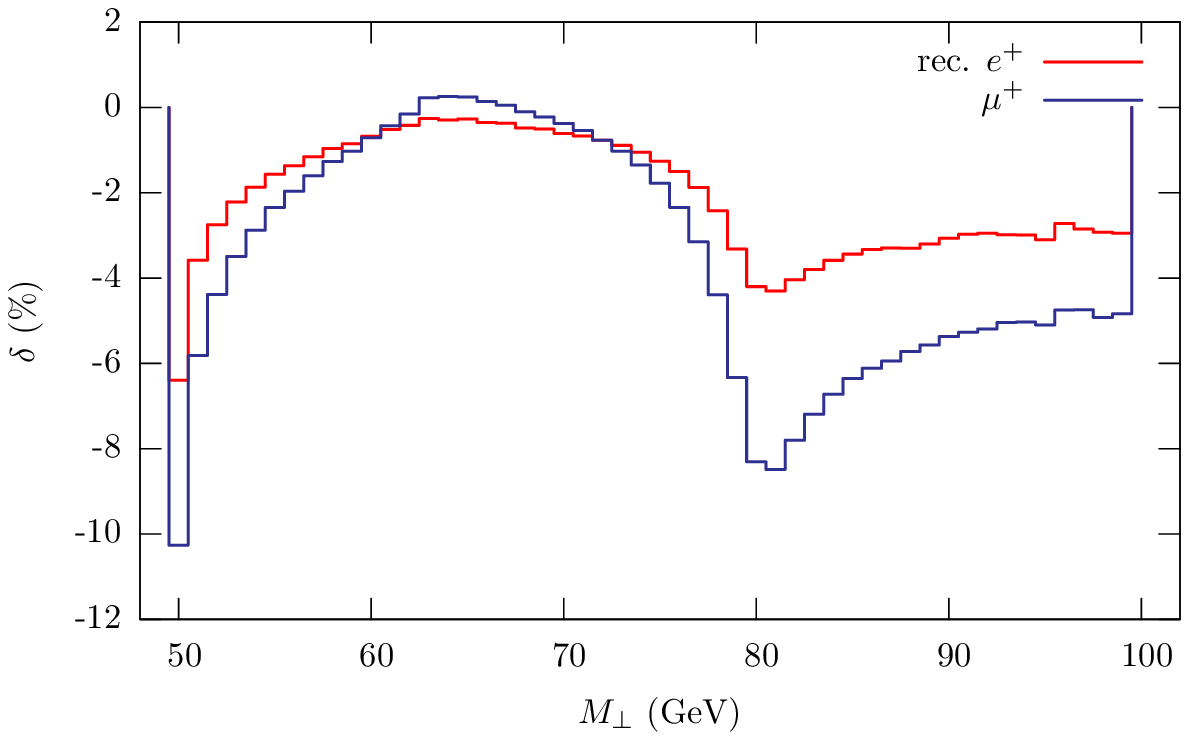}
\caption{Transverse mass distribution and \oa relative correction.}
\label{fig:1}       
\end{figure}
\begin{figure}
\centering
\includegraphics[height=35mm,angle=0]{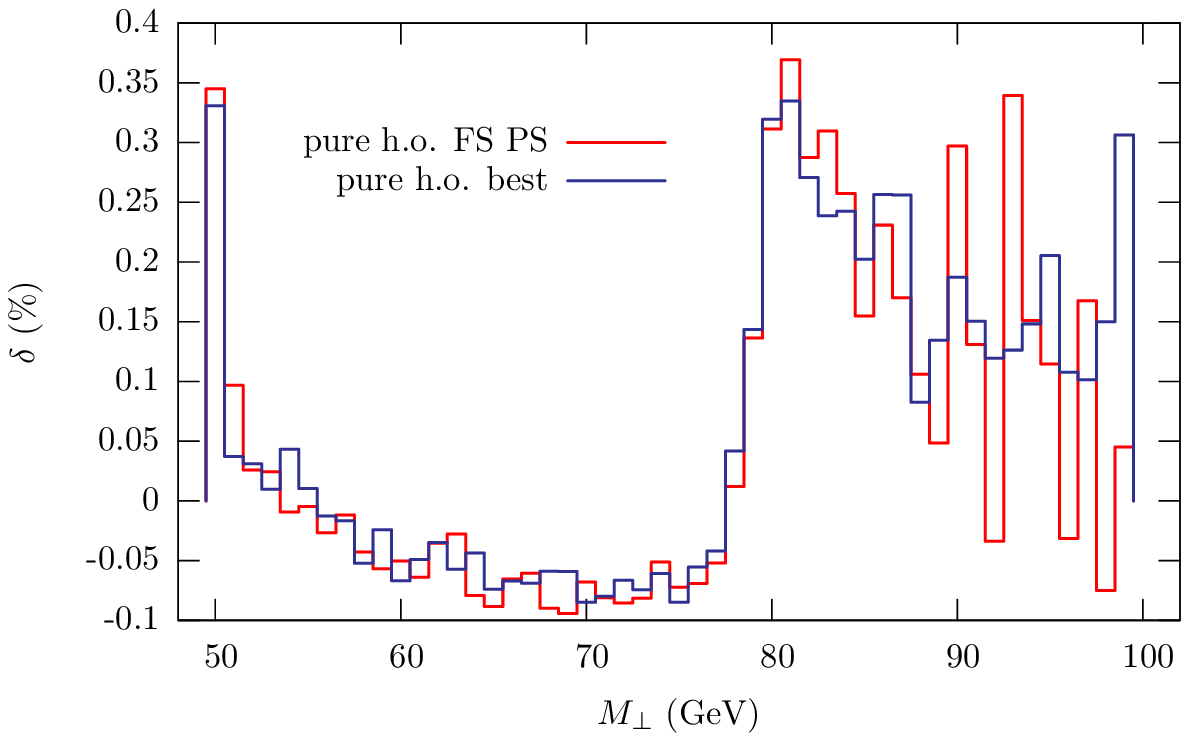}~
\includegraphics[height=35mm,angle=0]{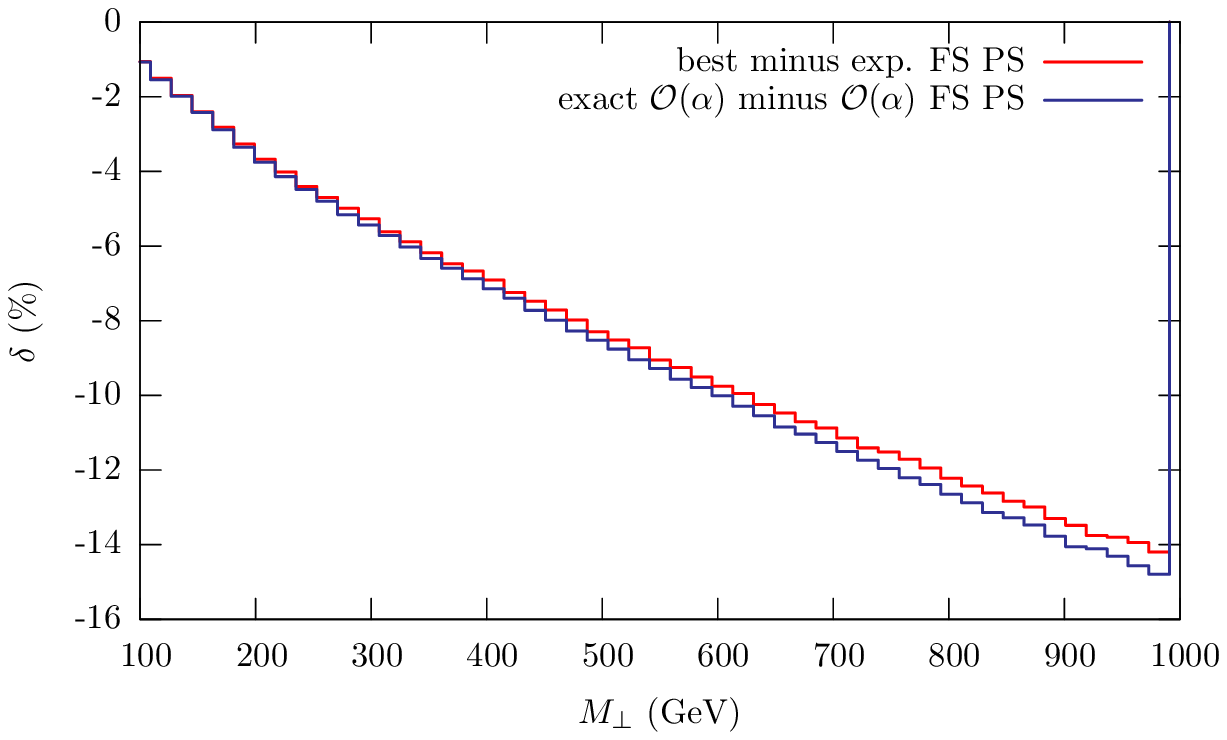}
\caption{Relative effect, with respect to the Born results, 
of the radiative corrections to the transverse mass distribution.
In the left panel the effect of the higher-order QED effects is
displayed. In the right panel the large negative corrections
due to the \oa EW Sudoakov logs is shown. }
\label{fig:2}       
\end{figure}
The generator
$\tt WGRAD$~\cite{BW} includes the exact \oa EW corrections.
The latter have been
shown to induce a shift on the value of $\mw$,
extracted at the Tevatron
from the study of the transverse mass distribution (see figure
\ref{fig:1}
\footnote{All the figures have been obtained with the new version
of  $\tt HORACE $ \cite{program} }),
of about 160 MeV in the muon channel~\cite{CDF},
mostly due to final-state QED radiation.
In view of the very high experimental precision foreseen at the LHC
($\Delta\mw\sim 15$~MeV),
final-state higher-order (beyond \oab) QED corrections may induce a
significant shift, as shown in ref.~\cite{CMNT} (see figure
\ref{fig:2}, left panel).
Some event generators
can account also for multiple-photon radiation: in the published
version of $\tt HORACE$~\cite{CMNT} final-state QED radiation was
simulated by means of a QED Parton Shower; the generator 
$\tt WINHAC$~\cite{WINHAC} uses the Yennie-Frautchi-Suura
formalism to
exponentiate final-state-like EW \oa corrections; finally, the
standard tool $\tt PHOTOS$~\cite{PHOTOS} can be used to describe QED
radiation in the $W$ decay.
A detailed series of tuned comparisons between different EW Monte
Carlo generators have been done in ref.~\cite{LesHouches},
in order to check the reliability of different numerical predictions,
with \oa accuracy,
in a given setup of input parameters and cuts.
\begin{figure}
\centering
\includegraphics[height=4cm]{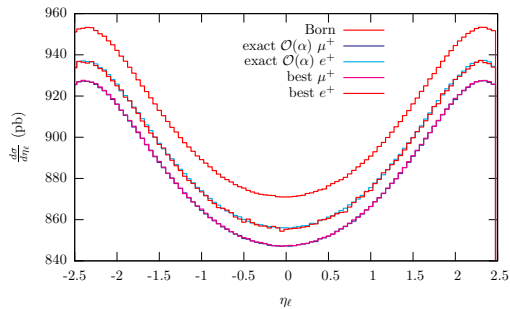}
\caption{Charged lepton pseudo-rapidity distribution. From top to
  bottom, Born, corrected electron, corrected muon distributions.}
\label{fig:3}       
\end{figure}

\noindent
{\bf The new HORACE version.}
A precision electroweak calculation of the charged current Drell-Yan process
has been recently completed \cite{CMNV}
and includes the exact \oa electroweak matrix elements and
leading-logarithmic QED higher-order corrections in the
Parton Shower approach.
The calculation is implemented in the new version of the 
Monte Carlo event generator $\tt HORACE$ \cite{program},
which combines, in a unique tool,
the good features of the QED Parton-Shower approach with the 
additional effects
present in the exact \oa EW calculation.
In figure \ref{fig:2} (right panel) it is shown the increasing size of
the EW \oa corrections, mostly due to the EW Sudakov logs.
They are important above the $W$ resonance, for a precise measurement
of the $W$ decay width and in the large transverse mass tail, to give
a precise estimate of the background to new gauge boson searches.
In figure \ref{fig:3} the effect of the \oa
and higher order QED corrections on the charged lepton pseudorapidity
distribution is presented. The latter is relevant to give a precise
determination of the detector acceptance, which in turn is a key
ingredient in the luminosity monitoring of the LHC.

\noindent
{\bf Perspectives.}
It is important, in view of the precision aimed in the $W$ mass
measurement, in the luminosity monitoring and in the new physics
searches,
to study the interplay
between QCD and EW corrections (a first analysis in this direction can
be found in ref. \cite{CY}).
Another possible upgrade of $\tt HORACE$ will be achieved with the
introduction of the exact EW \oa corrections also in the neutral
current channel.

%
%

%
%



\printindex
\end{document}